\documentclass[aps,pra,twocolumn,showpacs]{revtex4}

\usepackage{amsmath}
\usepackage{graphicx}

\allowdisplaybreaks

\begin{document}

\title{Dynamically turning off interactions in a two component condensate}
\author{D.~Jaksch, J.I.~Cirac, and P.~Zoller} 

\affiliation{Institut f\"{u}r
Theoretische Physik, Universit\"at Innsbruck, A--6020 Innsbruck,
Austria.} 

\begin{abstract}
We propose a mechanism to change the interaction strengths of a two
component condensate. It is shown that the application of $\pi/2$
pulses allows us to alter the effective interspecies interaction 
strength as well as the effective interaction strength between 
particles of the same kind. This mechanism provides a simple method
to transform spatially stable condensates into unstable once and vice
versa. It also provides a means to store a squeezed spin state by turning
off the interaction for the internal states and thus allows to gain 
control over many body entangled states.
\end{abstract}

\maketitle

\section{Introduction}
\label{Introduction}

The experimental achievement of atomic Bose-Einstein condensation (BEC) 
\cite{Cornell,Ketterle,General,Stringari} has stimulated extensive 
theoretical and experimental studies in this area. One of the most impressive 
examples of applications of BEC is to use condensates with internal degrees 
of freedom to generate quantum entanglement 
\cite{Sorensen,Pu,Duan,Duan2,Bigelow,Poulsen,Helmerson}, which is the 
essential ingredient for many quantum information protocols \cite{PhysWorld}. It 
has been shown that the coherent collisional interactions in BECs allow to
generate substantial many-particle entanglement in the spin degrees of freedom of
a two-component condensate \cite{Sorensen} during the free evolution of the 
condensates. The whole time evolution of the internal degrees of freedom 
is determined by the interaction strengths between the condensed particles. 
Thus it is desirable to control these interaction strengths by some external 
means since this opens the possibility to engineer many particle entangled states.

One possibility is to change the atomic interaction potential directly by applying an
external magnetic field which changes the scattering length. If one uses Feshbach 
resonances \cite{FeshbachT,FeshbachE,FeshbachE1} this method allows for considerable changes
in the interaction properties of BECs. In this paper we will propose another 
method to externally control the interaction strengths using $\pi/2$ 
pulses. This will not directly change the interatomic interaction potential but 
rather implement an {\em effective} Hamiltonian with an interaction strength depending 
on external parameters which can be adjusted easily. We will use this effective Hamiltonian
to study the influence of changing the interaction strengths on the external wave 
function. Furthermore, we will be able to control the time evolution of the 
internal degrees of freedom which are solely determined by the effective interaction 
strengths. In particular we will show how to turn off the Hamiltonian for the
internal states, i.e., the internal states will evolve like those in an ideal gas.

Let us briefly explain the basic idea for changing the interaction strengths of BECs 
using $\pi/2$ pulses: If we fix the spatial mode function of a two component condensate 
the dynamics of these two components is described by the Hamiltoinian
\cite{Sorensen,Duan2}
\begin{equation}
H_{\rm int}= \chi \tilde J_z^2,
\label{IntHameffIntro}
\end{equation}
where $\tilde J_z$ is the $z$-component of an angular momentum operator 
${\bf \tilde J}=\{\tilde J_x,\tilde J_y, \tilde J_z\}$ given by
\begin{eqnarray}
\tilde J_x &=& \frac 1 2 (a^\dagger b + b^\dagger a), \nonumber \\
\tilde J_y &=& \frac i 2 (b^\dagger a - a^\dagger b), \nonumber \\
\tilde J_z &=& \frac 1 2 (a^\dagger a - b^\dagger b),
\label{AngularMomentum2mode}
\end{eqnarray}
where $a$ ($b$) are bosonic destruction operators for particles in internal state
$1$ ($2$) with a spatial mode function $\varphi_{1(2)}$. The parameter $\chi$ is 
determined by the interaction properties of the two-component condensate. We apply 
$\pi/2$ Raman-laser or microwave pulses to the condensate which rotate the spin 
around the $x$ or $y$ axis by an angle $\pi/2$ depending on the phase 
of the pulse as we will show in Sec.~\ref{TwoModeApprox}. The idea is to 
use a sequence of $\pi/2$ pulses which rotates $H_{\rm int}$ to create contributions 
to the Hamiltonian given by $\chi J_x^2$ and $\chi J_y^2$ for a time $\zeta \; \delta t$
while the system evolves with $\chi J_z^2$ for a time $\delta t$.

Time averaging leads to an effective Hamiltonian which can be written as 
\begin{equation}
\tilde H^{\rm eff}_{\rm int} = \frac{\chi \tilde J_z^2 +\zeta \chi (\tilde J_y^2+\tilde J_x^2)}
{1+ 2 \zeta} \equiv \chi \frac{(1-\zeta) \tilde J_z^2 + \zeta {\bf \tilde J}^2}{1 + 2 \zeta}.
\label{ResIntHameffIntro}
\end{equation}
The operator ${\bf \tilde J}^2$  is a constant of motion and the Hamiltonian therefore is 
equivalent to $H^{\rm eff}_{\rm int}=\tilde \chi \tilde J_z^2$ up to a (time dependent) global 
phase. Thus, the application of the $\pi/2$ pulses effectively leads to a change in the 
interaction parameters from $\chi$ to $\tilde \chi=\chi (1-\zeta)/(1+2 \zeta)$. For $\zeta=1$ 
we find the effective Hamiltonian $H^{\rm eff}_{\rm int}$ to vanish. The internal states 
then evolve like those in a non-interacting gas.

The paper is organized as follows. In Sec.~\ref{ModelSec} we will introduce the
model. After writing down the Hamiltonian of a two component condensate interacting with 
a classical laser or microwave field we will define a specific series of pulses applied 
to the condensate. Then we will calculate the time averaged Hamiltonian and determine the
dependence of the effective interaction strengths on parameters of the external field. In 
Sec.~\ref{ApplSec} we will study possible applications of this effective Hamiltonian. We will 
investigate the influence of changing the interaction strength on the spatial wave functions 
of the condensate as well as on the evolution of the internal atomic degrees of freedom. 
Sec.~\ref{ImpSec} is devoted to the discussion of approximations and possible imperfections 
in our model. We conclude in Sec.~\ref{ConSec} with a discussion and summary of our results.

\section{Model}
\label{ModelSec}

In this section we start with the Hamiltonian of a two component
condensate interacting with an external field. We investigate the effect of 
$\pi/2$ pulses on the condensate and specify a specific series of pulses. 
We show that this specific choice of pulses effectively leads to a change 
in the interaction strengths of the condensate.

\subsection{Hamiltonian}

\begin{figure}[tbp]
\begin{center}
\includegraphics{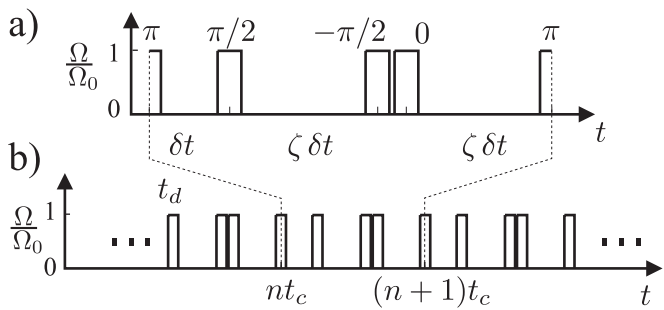}
\end{center}
\caption{Schematic plot of the sequence of laser pulses $\Omega/\Omega_0$ against time $t$ in 
arbitrary units. a) Sequence of $\pi/2$ pulses $\Omega(t)$ applied to the two component condensate 
between $n t_c$ and $(n+1) t_c$. The angle $\alpha$ of the $\pi/2$ pulses is shown above the 
corresponding pulse. Graph b) shows the whole series of pulses which consists of $\pi/2$ pulse 
sequences shown in a).
\label{fig1}}
\end{figure}

We consider a two component BEC consisting of $N$ atoms in different atomic hyperfine levels 
1 and 2 coupled by a time $t$ dependent classical field with Rabi frequency $\Omega(t)$ (internal 
Josephson effect \cite{JJ1,JJ2,JJ3,Ho}). The classical field can be realized by either 
a Raman-laser or by a microwave field applied to the condensate. We assume $\Omega$ to be ${\mathbf x}$ 
independent, i.e.~there is negligible momentum transfer to the condensates due to the interaction
with the classical field. The Hamiltonian of this system is given by $H=H_{\rm BEC}+H_{L}$, where 
$H_{\rm BEC}=H_1({\bf u})+H_2({\bf u})$ and $H_L$ describe the two component condensate and the 
interaction of external field and condensate, respectively \cite{Ketterle1,LYou,real}. 
These terms are given by (with $\hbar=1$)
\begin{subequations}
\begin{align}
H_{k}({\bf u})=& \int \! d^3{\mathbf x} \; \psi^\dagger_{k} \left(
-\frac{\nabla^{2}}{2m}
+V_k+\sum_{l}\frac{u_{kl}}{2} \psi^\dagger_{l} \psi_{l} \right) \psi_{k},\\
H_{L}=& \int \! d^3{\mathbf x} \; \left[\frac{\Omega(t)}{2} \psi_1^\dagger \psi_2 
+ {\rm h.c.} \right].
\end{align} \label{H}
\end{subequations}
Here, $\{k,l\} \in \{1,2\}$ and $\psi_{k} \equiv \psi_{k}({\mathbf x})$ is a
bosonic field operator which annihilates a particle at position ${\mathbf x}$
in hyperfine state $|k\rangle$. The trapping potential for particles in 
state $k$ is denoted by $V_k \equiv V_k({\mathbf x})$ and the mass of the atoms 
is $m$. The interaction strengths are given by ${\bf u}=\{u_{11},u_{22},u_{12}\}$ 
where for simplicity we assume $u_{11}=u_{22} \equiv u$ for the interaction between
atoms in the same internal state, and $u_{12} \neq u$ as is the case in Na \cite{real}.
Furthermore we assume the trapping potential for the different internal states to
be equal, i.e., $V_1=V_2 \equiv V$. We denote the first excitation energy of $V$ by 
$\omega$ and the size of the single particle ground state in the potential $V$ by $a_0$.

$H_L$ describes the interaction of the external field with the condensate. We assume 
that this external field is used to apply a sequence of $\pi/2$ pulses to the condensates.
While it is turned on the Rabi frequency is constant $\Omega(t)=\Omega_0$.
The duration of the $\pi/2$ pulses is thus given by $t_d=\pi/2 \Omega_0$ which is
assumed to be much shorter than the time scale determined by the evolution of
the condensates due to $H_k$, i.e., $\Omega_0 \gg \omega$, and $\Omega_0 \gg N u /a_0^3$
\cite{IntTimeScale}. Next we investigate the time evolution of the system while the external
field is turned on. 

\subsubsection{Interaction with the external field}

While $H_L$ is turned on, i.e., $\Omega \neq 0$ it is the dominant part of the Hamiltonian $H$ 
and we neglect contributions of $H_k$. A $\pi/2$ pulse is characterized by
\begin{equation}
\Omega(t)=|\Omega(t)| e^{i \alpha}, \qquad \int_{-\infty}^{\infty} |\Omega(t)| \; dt = \pi/2,
\label{pi2cond}
\end{equation}
with a phase $\alpha$ and it implements the following time evolution
for the bosonic field operators in the Heisenberg picture
\begin{equation}
U^\dagger_\alpha \left( \begin{array}{c} 
\psi_1 \\
\psi_2
\end{array} \right)
U_\alpha = \frac{1}{\sqrt{2}} 
\left( \begin{array}{cc} 
1 & -i e^{i \alpha} \\
-i e^{-i \alpha} & 1 
\end{array} \right) \left( \begin{array}{c} 
\psi_1 \\
\psi_2
\end{array} \right).
\label{pi2pulses}
\end{equation}
The inverse transformation is given by $U_\alpha^{\dagger}=U_\alpha^{-1}=U_{\alpha+\pi}$.

\subsubsection{Spin operators}
\label{SpinOp}

To get an intuitive picture of the effect of $\pi/2$ pulses on the condensate we define 
the spin operator ${\bf J}=\{J_x,J_y,J_z\}$ by
\begin{eqnarray}
J_x &=& \frac 1 2 \int \! d^3{\mathbf x} \; (\psi_1^\dagger \psi_{2} + \psi_{2}^\dagger \psi_1), \nonumber \\
J_y &=& \frac i 2 \int \! d^3{\mathbf x} \; (\psi_{2}^\dagger \psi_1 - \psi_1^\dagger \psi_{2}), \nonumber \\
J_z &=& \frac 1 2 \int \! d^3{\mathbf x} \; (\psi_1^\dagger \psi_1 - \psi_{2}^\dagger \psi_{2}).
\label{AngularMomentum}
\end{eqnarray}
In the Heisenberg picture these operators are transformed by the $\pi/2$ pulses according to
\begin{equation}
U_\pi J_z  U_0 = J_y, \qquad U_{-\pi/2} J_z  U_{\pi/2} = J_x.
\end{equation}
Thus the application of $U_0, (U_{\pi/2})$ to the condensate corresponds to a rotation of the 
spin $\bf J$ around the $x (y)$ axis by an angle $\pi/2$, respectively. 

\subsubsection{Series of $\pi/2$ pulses}
\label{SeriesLaserpulses}

We want to consider a specific sequence of pulses applied repeatedly to the condensates.
One sequence of pulses is shown in Fig.~\ref{fig1}a while Fig.~\ref{fig1}b shows the whole 
series of pulses. As can be seen from Fig.~\ref{fig1}a the condensate first evolves freely for 
a time $\delta t$. Then a $\pi/2$ pulse rotates the spin instantaneously around the $y$ axis by an 
angle $\pi/2$ and back after a time $\zeta \; \delta t$. Immediately afterwards we rotate the 
spin by $\pi/2$ around the $x$ axis and then back after time $\zeta \; \delta t$. This sequence 
of $\pi/2$ pulses is repeated as shown in Fig.~\ref{fig1}b. Each sequence takes a time 
$t_c=(1+2 \zeta) \delta t$, neglecting the time $t_d$ needed to apply a pulse.

There are thus four time scales in this model: (i) $t_d$ the duration of a $\pi/2$ pulse,
(ii) $t_c$ which is the time needed for applying a sequence of $\pi/2$ pulses, (iii)
$t_{\rm BEC}=1/\omega$ which determines the time scale of the free evolution of the condensates
and (iv) $t_{\rm int}=a_0^3/ u N$ which is the time scale set by the interactions between the
particles \cite{IntTimeScale}. These four time scales are assumed to satisfy the relation 
$t_d \ll t_c \ll t_{\rm BEC} \approx t_{\rm int}$ which can easily be achieved by an appropriate 
choice of external parameters.

\subsubsection{Time averaged Hamiltonian}

\begin{figure}[tbp]
\begin{center}
\includegraphics{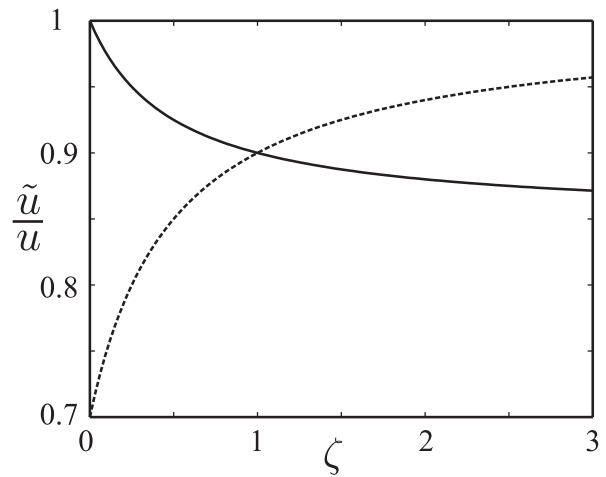}
\end{center}
\caption{Interaction strength $\tilde u$ (solid curve) for particles in the same hyperfine level
and $\tilde u_{12}$ (dashed curve) for particles in different hyperfine levels as a function 
of $\zeta$. For $\zeta=0$ we assumed that $\tilde u_{12}=u_{12}=0.7 u=0.7 \tilde u$.
\label{fig2}}
\end{figure}

We now want to study the time evolution of the condensates when the pulses specified in
the previous section \ref{SeriesLaserpulses} are applied to the system. The time evolution 
operator ${\cal U}^M$ at time $t=M t_c$, i.e., after applying $M$ pulse sequences is given by
\begin{equation}
{\cal U}^M= \prod_{l=1}^M {\cal U},
\end{equation}
where
\begin{eqnarray}
{\cal U}&=&U_{\pi} e^{-i H_{\rm BEC} \zeta \; \delta t} U_0 \times \nonumber \\
&&\quad U_{-\pi/2} e^{-i H_{\rm BEC} \zeta \; \delta t} U_{\pi/2} e^{-i H_{\rm BEC} \delta t} 
\nonumber \\
&\equiv& e^{-i H_{\rm eff} t_c}.
\end{eqnarray}
To first order in $t_c$ we find for the effective Hamiltonian
\begin{eqnarray}
H_{\rm eff}&=&\frac{H_{\rm BEC}+ \zeta ( U_{\pi/2} H_{\rm BEC} U_{-\pi/2}+ U_0 H_{\rm BEC} 
U_{\pi})}{1+ 2 \zeta} \nonumber \\
&=& H_1({\bf \tilde u})+H_{2}({\bf \tilde u}),
\label{HamilEff}
\end{eqnarray}
where ${\bf \tilde u}=\{\tilde u_{11},\tilde u_{22}, \tilde u_{12}\}$ with
$\tilde u_{11}=\tilde u_{22} \equiv \tilde u$ and
\begin{eqnarray}
\tilde u &=& \frac{u+(u+u_{12}) \zeta}{1+ 2 \zeta}, \nonumber \\
\tilde u_{12} &=& \frac{u_{12}+2 u \zeta}{1+ 2 \zeta}.
\end{eqnarray}
The effective interaction strengths $\tilde u$ and $\tilde u_{12}$ appearing in the time 
averaged Hamiltonian $H_{\rm eff}$ depend on the parameter $\zeta$. Experimentally this 
parameter can easily be changed by adjusting laser or microwave parameters.
In Fig.~\ref{fig2} we show the dependence of the interaction strength on the 
parameter $\zeta$. The situation $\zeta=0$ corresponds to the case of applying 
no pulses. We have assumed $u>u_{12}$, i.e., a spatially stable two component 
condensate. By increasing $\zeta$ we find $\tilde u$ decreases and $\tilde u_{12}$
increases. At $\zeta=1$ the two effective interaction strengths are crossing. This
situation corresponds to the case where the internal Hamiltonian $H_{\rm int}^{\rm eff}=0$
as discussed in Sec.~\ref{Introduction}.

Qualitatively we can understand how an effective change in the interaction strengths arises
by looking at the time evolution of a condensate with initially all the particles in state 
$|k\rangle$. If no laser pulses are applied to the condensate the phase accumulated due to 
the interaction during a time $\zeta \delta t$ is proportional to $u \zeta \delta t$. However, if
a $\pi/2$ pulse is applied the condensate is put into a superposition state of the internal 
states $|1\rangle$ and $|2\rangle$ according to Eq.~(\ref{pi2pulses}). In this case the phase 
accumulated during a time $\zeta \delta t$ is proportional $(u+u_{12}) \zeta \delta t/2$. 
Then a second $\pi/2$ pulse brings the condensate back to the internal state $|k\rangle$. 
Effectively the two $\pi/2$ pulses thus lead to a change of the interaction strength.

\subsection{Two mode approximation}
\label{TwoModeApprox}
In this section we use the two mode approximation for deriving the coupled
Gross-Pitaevskii equations (GPE) for the condensate. Then we investigate the time
evolution of the internal atomic degrees of freedom. 

\subsubsection{External degrees of freedom}
We assume that the condensate can be described using only
one spatial mode function for each component, i.e.,
\begin{equation}
\psi_1({\mathbf x},t)=a \; \varphi_1({\mathbf x},t), \quad \text{and} 
\quad \psi_2({\mathbf x},t)=b \; \varphi_2({\mathbf x},t),
\label{TwoModeAnsatz}
\end{equation}
with $\varphi_k({\mathbf x},t)$ 
the spatial wave function of condensate $k$ and $a(b)$ bosonic annihilation
operators for particles in condensate 1(2). We put this ansatz into the Hamiltonian Eq.~(\ref{HamilEff}), 
assume the state of the condensate to be 
\begin{equation}
|\psi\rangle=\frac{(a^\dagger)^{N_1} (b^\dagger)^{N_2}}{\sqrt{N_1! N_2!}} |0\rangle,
\end{equation}
with $N_1$ particles in condensate $1$, $N_2$ particles in condensate $2$ and 
$|0\rangle$ being the vacuum state.
Minimizing the expression
\begin{equation}
\langle \psi | i \frac{\partial}{\partial t} - H_{\rm eff} | \psi \rangle,
\label{minans}
\end{equation}
with respect to the wave functions $\varphi_{1,2}({\mathbf x})$ we find the coupled 
GPE equations
\begin{eqnarray}
i \frac{\partial \varphi_1}{\partial t}&=&\left(-\frac{\nabla^{2}}{2m}
+V+\tilde u \; N_1 |\varphi_1|^2+ \tilde u_{12} \; N_2 |\varphi_2|^2 \right) 
\varphi_1, \nonumber \\
i \frac{\partial \varphi_2}{\partial t}&=&\left(-\frac{\nabla^{2}}{2m}
+V+\tilde u \; N_2|\varphi_2|^2+ \tilde u_{12} \; N_1|\varphi_1|^2 \right) 
\varphi_2, \nonumber \\ \label{GPEcoupled}
\end{eqnarray}
where $\varphi_{1,2} \equiv \varphi_{1,2}({\mathbf x},t)$ and $\tilde u$ and
$\tilde u_{12}$ are time dependent.

\subsubsection{Internal degrees of freedom}

We want to simplify the model further by assuming $N_1-N_2=n$ being of order 
${\cal O}(\sqrt{N})$ and set $\varphi_1 = \varphi_2 \equiv \varphi$ in the ansatz 
for the bosonic field operators. Minimizing the terms of order ${\cal O}(N)$ in 
expression Eq.~(\ref{minans}) we find the GPE
\begin{equation}
i \frac{\partial \varphi}{\partial t}=\left(-\frac{\nabla^{2}}{2m}
+V+N \frac{\tilde u +\tilde u_{12}}{2} |\varphi|^2 \right) \varphi,
\end{equation}
for the wave function. If $\varphi$ fulfills the above GPE the terms of 
order ${\cal O}(\sqrt{N})$ vanish in Eq.~(\ref{minans}) and the time 
evolution of the internal atomic degrees of freedom is given by the Hamiltonian
(up to a global phase)
\begin{equation}
H^{\rm eff}_{\rm int}= \tilde \chi \tilde J_z^2,
\label{IntHameff}
\end{equation}
where 
\begin{equation}
\tilde \chi=\chi \frac{1-\zeta}{1+2 \zeta}, \quad
\text{with} \quad 
\chi=(u - u_{12}) \int \! d^3{\mathbf x} \; \left| \varphi \right|^4.
\label{ChiDef}
\end{equation}
At $\zeta=1$ the parameter $\tilde \chi=0$. There will thus be no dynamics of 
the internal degrees of freedom even for $N_1 \neq N_2$. The internal spin operator 
${\bf \tilde J}$ is obtained from $\bf J$ by using ansatz Eq.~(\ref{TwoModeAnsatz}) 
and is explicitely given in Eq.~(\ref{AngularMomentum2mode}). A physical interpretation 
of the evolution of the internal atomic states was already given in the introduction 
Sec.~\ref{Introduction}.

Putting the two mode ansatz Eq.~(\ref{TwoModeAnsatz}) into the time evolution 
operator ${\cal U}^M$ we find
\begin{eqnarray}
{\cal U}^M_{\rm int} &=& \prod_{l=1}^M \left[\tilde U_{\pi} e^{-i H_{\rm int} \zeta \; 
\delta t} \tilde U_0 \times \right. \nonumber \\
&& \left. \quad \tilde U_{-\pi/2} e^{-i H_{\rm int} \zeta \; \delta t} \tilde U_{\pi/2} 
e^{-i H_{\rm int} \delta t}\right],
\end{eqnarray} 
where $\tilde U_\alpha$ is obtained from $U_\alpha$ by replacing the bosonic field
operators $\psi_{1(2)}$with the corresponding annihilation operators $a$ ($b$).

\section{Applications}
\label{ApplSec}

In this section we study possible applications of changing the interaction
strength between the condensate discussed above. First we show how stable 
condensates can be destabilized and vice versa. Then we investigate how 
a squeezed state of a two component condensate can be preserved by turning
off the interactions for the internal states.

\subsection{(Un)stabilizing a two component condensate}

The properties of multi-component condensates such as spin domain formation have
been studied extensively both experimentally \cite{Ketterle1,Myatt,Hall,Stenger,Ketterle3} and
theoretically \cite{Ho,Ao,Ohberg,Law} in the last few years for constant interaction strengths. 
Here, we study the effect of varying $\zeta$ and
thus changing the interaction strengths between the condensed particles on the condensate 
wave functions $\varphi_{k}$ and the spatial stability. For $\tilde u > \tilde u_{12}$ two 
initially overlapping condensates should remain spatially stable while for $\tilde u < \tilde u_{12}$ 
they separate. 

\begin{figure}[tbp]
\begin{center}
\includegraphics{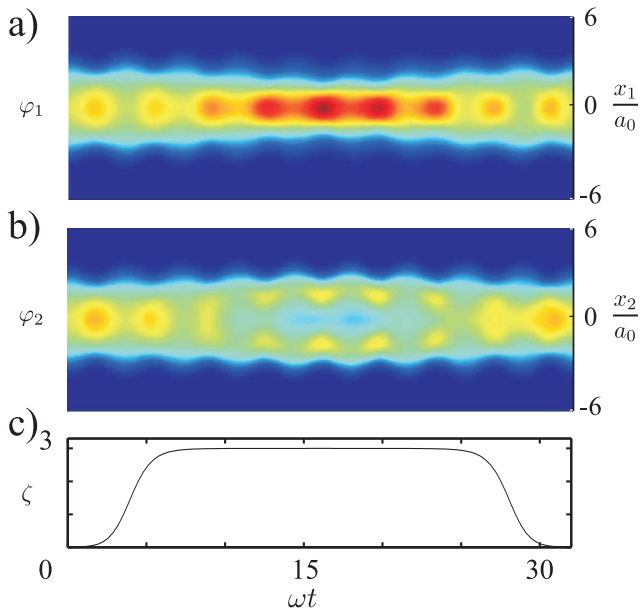}
\end{center}
\caption{Condensate wave functions $\varphi_{1[2]}$ as a function
of time and spatial coordinate $x_{1[2]}$ are shown in graphs a [b)]. 
Red regions indicate large condensate density, blue regions do not contain condensed particles. 
c) Parameter $\zeta$ as a function of time. The parameters are chosen $N_1=5200$, $N_2=4800$, 
trapping frequency $\omega=40$Hz, the interaction strength is assumed to be 
$2 u_{12}=u=3 \cdot 10^{-3} \omega a_0$. \label{fig3}}
\end{figure}

We solve numerically Eq.~(\ref{GPEcoupled}) in one spatial 
dimension. The trapping potential is assumed to be harmonic $V(x)=m \omega^2 x^2 /2$ with 
$\omega$ the trap frequency and $a_0=\sqrt{\hbar / m \omega}$ the ground state size. 
Figure \ref{fig3}a(b) shows the condensate wave functions $\varphi_{1(2)}$, respectively. We 
change $\zeta(t)$ as shown in Fig.~\ref{fig3}c. Initially $\zeta(0)=0$ and we assume that 
$u>u_{12}$, i.e., the two condensates are strongly overlapping. When $\zeta(t)>1$ the 
repulsion between atoms in different hyperfine states separates the condensates in space 
since then $u_{12}>u$ (as can be seen from Fig.~\ref{fig2}). As soon as $\zeta(t)$ becomes 
smaller than $1$ again the two condensates become overlapping again. Note that the time 
scale for the separation of the two condensates depends also on the imbalance of the 
condensate particle numbers $N_1-N_2$. For $N_1=N_2$ the two condensates do not separate 
for the parameters chosen in Fig.~\ref{fig3}.

\subsection{Spin squeezed condensate states}
Next we want to consider the behavior of the internal degrees of
freedom and show that by choosing $\zeta=1$ and thus making the
internal Hamiltonian $H_{\rm int}=0$ we can store a spin squeezed
state. 

\subsubsection{The squeezing parameter}

The entanglement properties of the atoms can be expressed in terms 
of the variances and expectation values of the angular momentum 
operators $\bf \tilde J$. Of particular interest is the squeezing
parameter $\xi^2$ defined by \cite{Sorensen}
\begin{equation}
\label{xi}
\xi^2 = \min_{{\bf n_{1,2,3}}} \frac{N (\Delta \tilde J_{{\bf n}_1})^2}
{\langle \tilde J_{{\bf n}_2} \rangle^2 +
\langle \tilde J_{{\bf n}_3}\rangle^2},
\end{equation}
where $\tilde J_{\bf n} \equiv {\bf n \cdot \tilde J}$ and 
the ${\bf n_{1,2,3}}$ are mutually orthogonal unit vectors.
If $\xi^2 < 1$ the state of the atoms is non--separable 
(i.e. entangled) as has been shown e.g.~in \cite{Sorensen}.
The parameter $\xi^2$ thus characterizes the atomic entanglement, and
states with $\xi^2 < 1$ are often referred to as ``spin squeezed 
states'' \cite{kitagawa}.

\subsubsection{Preserving a spin squeezed condensate state}

\begin{figure}[tbp]
\begin{center}
\includegraphics{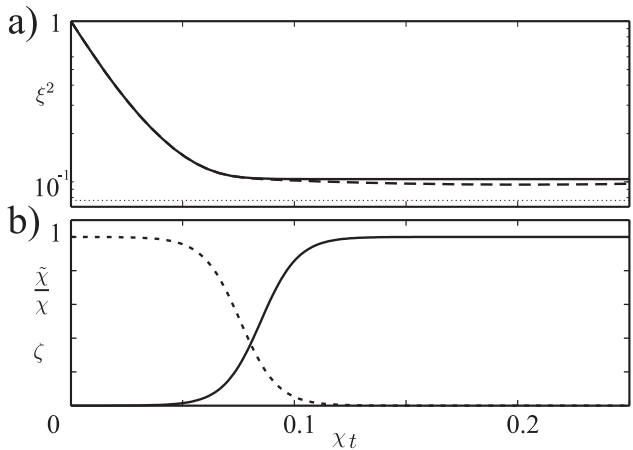}
\end{center}
\caption{a) Squeezing parameter $\xi^2$ as a function of time. The dotted line shows
the minimum squeezing parameter $\xi^2_m$ achievable by one axis squeezing as defined
in the text. The solid curve shows the squeezing parameter obtained by solving the
Schr{\"o}dinger equation using the effective Hamiltonian $H^{\rm eff}_{\rm int}$ defined in
Eq.~(\ref{IntHameff}). The dashed curve shows the squeezing parameter obtained from
the time evolution operator ${\cal U}^M_{\rm int}$ with $\chi t_c=5 \cdot 10^{-3}$. b) Parameter 
$\zeta$ (solid curve) and resulting relative interaction strength $\chi$ (dashed 
curve) as functions of time. The numerical calculation was done for $N=50$.
\label{fig4}}
\end{figure}

We assume an initial state of the form
\begin{equation}
|\psi\rangle=\frac{(a^\dagger+b^\dagger)^N}{\sqrt{N!}} |0\rangle
\end{equation}
created by applying a $\pi/2$ pulse with $\alpha=\pi/2$ to a condensate
of particles in internal state $|1\rangle$. The evolution of this initial
state according to the Hamiltonian $H_{\rm int}^{\rm eff}$ with constant 
$\tilde \chi$ has been studied extensively in \cite{Sorensen} and leads to
one axis squeezing as defined in \cite{kitagawa}. Initially $\xi^2=1$ and 
is then rapidly reduced. After reaching a minimum value the entanglement parameter
$\xi^2$ increases again. Our aim is to control the interaction parameter
$\tilde \chi$ such that after $\xi^2$ has reached its minimum value further 
evolution of the system is suppressed. In Fig.~\ref{fig4} we show a comparison 
of the time evolution with the time evolution operator ${\cal U}^M_{\rm int}$, 
and the effective Hamiltonian $H^{\rm eff}_{\rm int}$. The squeezing parameter $\xi^2(t)$ 
is shown in Fig.~\ref{fig4}a and the time dependence of $\zeta(t)$ is shown in 
Fig.~\ref{fig4}b. As soon as $\xi^2$ has reached its minimum value $\zeta$ goes rapidly towards one and 
thus prohibits further evolution of $\xi^2$. The squeezing parameter $\xi^2$ remains 
at its minimum value which is close to the minimum value $\xi^2_m=(3/N)^{2/3}/2$ 
that can be reached by one axis squeezing \cite{kitagawa}. Note that the minimum squeezing
parameter that is reached by the evolution according to ${\cal U}^M_{\rm int}$ is smaller
than expected from the Hamiltonian $H_{\rm eff}$ as long as $t_c \ll t_{\rm BEC}$
is fulfilled (cf. Fig.~\ref{fig4}). We find, however, that this difference is
always very small and vanishes if we further decrease $t_c$ compared to the 
value used in Fig.~\ref{fig4}.

\section{Discussion}
\label{ImpSec}

There are two different kinds of approximations in our scheme. First, the
Hamiltonian $H_{\rm eff}$ we use is time averaged over the duration of a
sequence of pulses $t_c$ and second, we use a two mode description of the
two component condensate for describing the dynamics of the internal states.
We will discuss these two approximations separately since they are independent
of each other. Also, experimentally it is not possible exactly realize $\pi/2$ 
pulses. Therefore, we will also discuss the influence of imperfections in
the $\pi/2$ pulses.

\subsection{Approximations}

\subsubsection{Time averaging}

While $H_L$ is turned on we neglect the free time evolution of the condensate
due to $H_{\rm BEC}$ completely. Typically the time evolution due to the applied
$\pi/2$ pulses will take place on a time scale $t_d$ of a few ns while the typical 
time scale for the free evolution of the condensate $t_{\rm BEC}$ is on the order
of ms. The neglect of the free evolution of the condensates during a pulse
will lead to an error of the order of $t_d/t_{\rm BEC} \approx 10^{-4}$ and is thus
well justified. The second step in calculating the effective Hamiltonian is to 
average over one sequence of pulses. This will typically lead to an error on
the order of $t_c/t_{\rm BEC}$. In Fig.~\ref{fig4} we compare the time evolution
according to the time averaged Hamiltonian with the time evolution given by ${\cal U}^M_{\rm int}$
for $t_c \chi=5 \times 10^-3$ and find a very small deviation between the two results.

\subsubsection{Two mode approximation}

The form of the effective Hamiltonian $H_{\rm eff}$ is equivalent to the standard
form of the Hamiltonian for two component BECs. Therefore we expect the same range
of validity for $H_{\rm eff}$ as for the original Hamiltonian $H$. This also applies
to the two mode approximation introduced in Sec.~\ref{TwoModeApprox}.

\subsection{Imperfect $\pi/2$ pulses}

\begin{figure}[tbp]
\begin{center}
\includegraphics{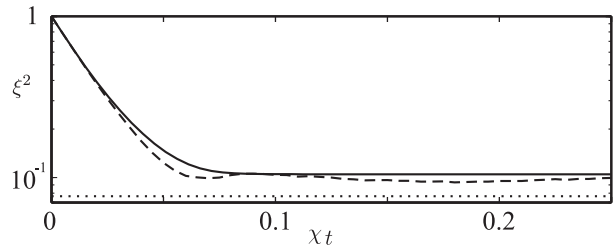}
\end{center}
\caption{Squeezing parameter $\xi^2$ as a function of time. The dotted line shows
the minimum squeezing parameter $\xi^2_m$ achievable by one axis squeezing as defined
in the text. The solid curve shows the squeezing parameter obtained by solving the
Schr{\"o}dinger equation using the effective Hamiltonian $H^{\rm eff}_{\rm int}$ defined in
Eq.~(\ref{IntHameff}). The dashed curve is obtained by calculating the time evolution
according to the time evolution operator ${\cal U}^M_{\rm int}$ with $\chi t_c=5 \cdot 10^{-3}$ and
a 1\% error in the intensity of the $\pi/2$ pulses. The ensemble average over $R=2000$ realizations
is shown. The other parameters are equal to those chosen in Fig.~\ref{fig4}.
\label{fig5}}
\end{figure}

It is experimentally possible to adjust the phase $\alpha$ of the $\pi/2$ pulses
very precisely, while it is much harder to exactly fulfill the integral condition 
in Eq.~(\ref{pi2cond}). Therefore we investigate the influence of a violation  of this 
condition on the time evolution of our system. We assume a random Gaussian error of 1\% in 
the value of the integral in Eq.~(\ref{pi2cond}) for each pulse applied to the condensate and
calculate the resulting time evolution in the two mode approximation. Figure \ref{fig5}
shows the result for the squeezing parameter averaged over $R=2000$ different realizations.
As can be seen from Fig.~\ref{fig5} an error in the duration and intensity of the pulses
does not lead to a qualitatively different behavior of the system. For some of the realizations
we obtain a smaller squeezing parameter $\xi^2$ than expected from one axis squeezing.
In this case the error in the $\pi/2$ pulses leads to some two axis squeezing which yields
a smaller squeezing parameter than pure one axis squeezing \cite{kitagawa}.

\section{Conclusions}
\label{ConSec} 

In this paper we have introduced a method to change the interaction strength of a two 
component condensate by $\pi/2$ pulses. We have shown that applying a specific series of
pulses to the condensate leads to an effective time averaged Hamiltonian which
is of the form of the original two component Hamiltonian with an interaction strength
depending on parameters of the external field.

As applications of this scheme we have proposed to use this Hamiltonian for turning a
stable condensate into an unstable one and vice versa. We have also shown that it is
possible to store a spin squeezed state of a condensate for, at least in principle, an
arbitrarily long time. 

Finally, we want to point out that the method to change the interaction
strengths of BECs discussed in this paper can experimentally be realized with current
technology. It is intended to serve as a tool to gain further insight into the properties
of BECs as well as to aim in engineering many particle entangled states.

\acknowledgments
We thank Lu-Ming Duan and F.~Schmidt-Kaler for stimulating discussions.
This work was supported by the Austrian Science Foundation 
(Projekt Nr. Z30-TPH, Wittgenstein-Preis and SFB ``Control 
and measurement of Coherent Quantum Systems'').


\begin{thebibliography}{99}
\bibitem{real}
This Hamiltonian can be realized by e.g.~using the hyperfine atomic levels
$|F=1,m_F\pm 1 \rangle$ of the Na $F=1$ manifold as described in \cite{Ketterle1}.
To avoid spin relaxation processes to the $|m_F=0\rangle$ hyperfine state an 
external field that shifts this level out of resonance can be used. The remaining
states also fulfill our assumption $u_{11}=u_{22}=u$ and $u_{12} \neq u$. Note that 
the last assumption does not hold for ${}^{87}$Rb atoms where $u_{12} \approx u$.
\bibitem{IntTimeScale}
Here we implicitly assume that $u_{11}$, $u_{22}$ and $u_{12}$ are of the same
order of magnitude.
\end{thebibliography}
\end{document}